%% file: ramsey_communities_v5.tex
\documentclass[reprint,aps,pre,amsmath,amssymb,superscriptaddress,showpacs]{revtex4-1}
\usepackage{graphicx}
\usepackage{hyperref}

\usepackage{color}
\usepackage{soul}

\newtheorem{definition}{Definition}

\begin{document}

\title{Emergence of network communities driven by local rules}

\author{Alexei Vazquez}
\email{alexei@nodeslinks.com}
\affiliation{Nodes \& Links Ltd, Salisbury House, Station Road, Cambridge, CB1 2LA, UK}

\begin{abstract}
Natural systems are modeled by networks with nodes and links. Often the nodes are segregated into communities with different connectivity patterns. Node heterogeneity such as political affiliation in social networks or biological function in gene networks are highlighted as key factors driving the segregation of nodes into communities. Here, by means of numerical simulations, I show that node heterogeneity is not a necessary requirement. To this end I introduce the Ramsey community number, $r_ \kappa$, the minimum graph size that warranties the emergence of network communities with almost certainty. Using the stochastic block model and Infomap methods for community detection, I show that networks generated by local rules have finite $r_ \kappa$ values while their randomized versions do not have emergent communities. I conjecture that network communities are an emergent property of networks evolving with local rules.
\end{abstract}

\maketitle

\section{Introduction}

Communities are a salient feature of real networks \cite{newman_communities_2006}. People segregate into social groups based on profession or political affiliation. Gene evolution is shaped by the underlying biological functions. These communities are seen as heterogeneous connectivity patterns between nodes of different types. With that assumption at hand, the research on network communities has focused on the nuances of inferring community structures \cite{radicchi_communities_2004, newman_communities_2006, fortunato_communities_2007, hofman_communities_2008, karrer_blockmodels_2011, peixoto2024networkreconstructionminimumdescription}. More recently, into the inference of network communities in networks with higher order structures \cite{vazquez09hg, contisciani_inference_2022}.

In 2001 Jin, Girvan and Newman (JGN) suggested a different perspective \cite{growing_social_networks_jin01}. They introduced a model of social networks dynamics with a triadic closure rule: pairs of individuals become friends at a rate proportional to how many mutual friends they have. By means of numerical simulations they demonstrated that this rule generates networks with communities,  even though all nodes are of the same type. Similar observations were made by Sol\'e and Valverde for the gene duplication and divergence model of protein-protein interaction networks \cite{sole_spontaneous_2007}. The induction of network communities by triadic closure was later consolidated by Bianconi {\em et al} \cite{communities_bianconi_2014} and Wharrie {\em et al} \cite{communities_wharrie_2019}. These observations open good questions: Could the segregation we observe in real networks be determined or initiated by a process of network evolution? Do we need triadic closure or would cycles of larger length induce the formation of network communities as well?

This type of questions belongs to the Ramsey theory branch of combinatorics \cite{graham1980ramsey}. Ramsey theory cares about the appearance of order in random structures. A typical question in Ramsey theory is how big should a random structure be such that a specific motif is present with almost certainty. With that in mind we ask the question: Given some network evolution rules, how big should the network be to observe network communities with almost certainty? That is the subject of this work. The Ramsey theory of community structure in dynamical networks.

The first obvious question is what is peculiar about the JGN model leading to the appearance of communities. Based on previous experience \cite{vazquez03local}, I conjecture it is the local nature of the network evolution. According the JGN rules, node pairs must share a common neighbor to make a new connection between them. Link creation is restricted to a local neighborhood. To support this conjecture I will analyze generative network models with local and non-local evolution rules.

The paper is organized as follows. In Sec. \ref{sec_methods} I describe the core network methods used throughout this work. Unless specified, the results are obtained using the stochastic block model with correction for the degree sequence. In Sec. \ref{sec_models} I introduce the network generative models and a preliminary analysis of their community structure. In Sec. \ref{sec:other} I investigate the use of different community methods: modularity, infomap and the stochastic block model without correction for the degree sequence. Then in Sec. \ref{sec_ramsey} I report the main result of this work, the definition of the {\em Ramsey community number} and the {\em emergent communities property}. I illustrate their application to characterize the emergence of communities in the studied generative network models. In Sec. \ref{sec:ba_revisited} I investigate generalizations of the Barab\'asi-Albert model allowing for a baseline attractiveness. These are generative models that do not specify a local structure in their evolution rules, but are characterized by a high clustering coefficient. In Sec. \ref{sec:cycles} I investigate the impact of the cycle length on the emergence of communities.  Finally, I make some general conclusions in Sec \ref{sec_conclusions}.

\section{Choice of community algorithms}
\label{sec_methods}

We need to discuss some technicalities before we start. The first one is about the method to determine the network communities. Throughout this work I will use the stochastic block model implemented in the software package graph-tool  (\verb|graph_tool.inference.minimize_blockmodel_dl|, with default parameters)  \cite{peixoto_graph-tool_2014}). This stochastic block model finds the community structure with the minimum description length \cite{peixoto2024networkreconstructionminimumdescription}. In that sense, it gives as output the optimal number of communities $\kappa$ and the partition of the nodes into communities. An important parameter is the correction for the network degree sequence, \verb| deg_corr: bool (optional, default: True)|. Some of the networks investigated have power law degree distributions, and therefore I use the degree-corrected version of the stochastic block model (default option). The need for this correction will become evident in Sec. \ref{sec:other}, where different community detection methods are investigated. I have not considered stochastic block models with correction for triadic closure \cite{triadic_peixoto_2022}, because of two reasons. First, that is part of the hypothesis being tested, that local rules induce community structures. Second, there are local rules that generate networks with no triangles but do have communities. In the latter context correcting for triadic closure makes no difference.

I have tried two other methods for community detection. One is the modularity optimization with the Louvain method \cite{communities_blondel_2008} implementated in \verb| networkx.louvain_communities | \cite{networkx}. The other one is the Infomap method implemented in the package with the same name \cite{mapequation2025software}. The latter minimizes the description length of random walkers on the network.

The second technical point is about the statistical description of network communities. For every generative model, network size and model parameters, I will inspect 1000 instances. From those instances I will then estimate the probability $P_ \kappa = {\rm Prob}(\kappa\geq 2)$ of having a community structure and the average number of communities $\langle \kappa\rangle$.

Finally, we need a rewiring algorithm preserving the nodes degrees. I use the standard configuration model implemented with \verb|graph_tool.generation.random_rewire| with default parameters. The configuration algorithm rewires the network links preserving the degree distribution \cite{configuration_method_park_2004}.

The core methods to generate the data presented here can be found at \href{https://github.com/av2atgh/ramsey_netcom}{github.com/av2atgh/ramsey\_netcom}.

Finally, I will use the greek letter $\kappa$ to indicate network communities, based on the greek word for community $\kappa o \iota \nu \acute{o} \tau \eta \tau \alpha$.

\begin{figure}
\includegraphics[width=3.3in]{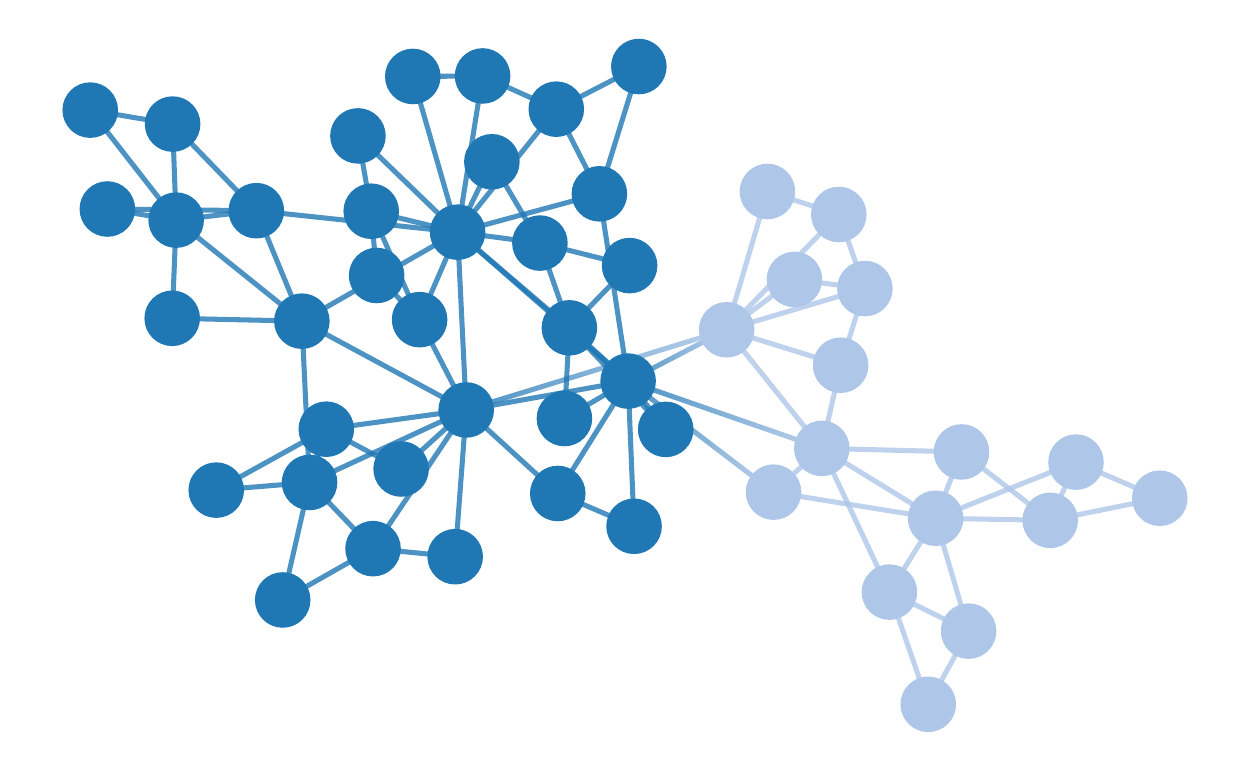}
\caption{An instance of LS$(n=50,\ d=1)$. The coloring represents the community structure inferred with the stochastic block model with degree sequence correction.}
\label{fig_ls1_example}
\end{figure}

\begin{figure}
\includegraphics[width=3.4in]{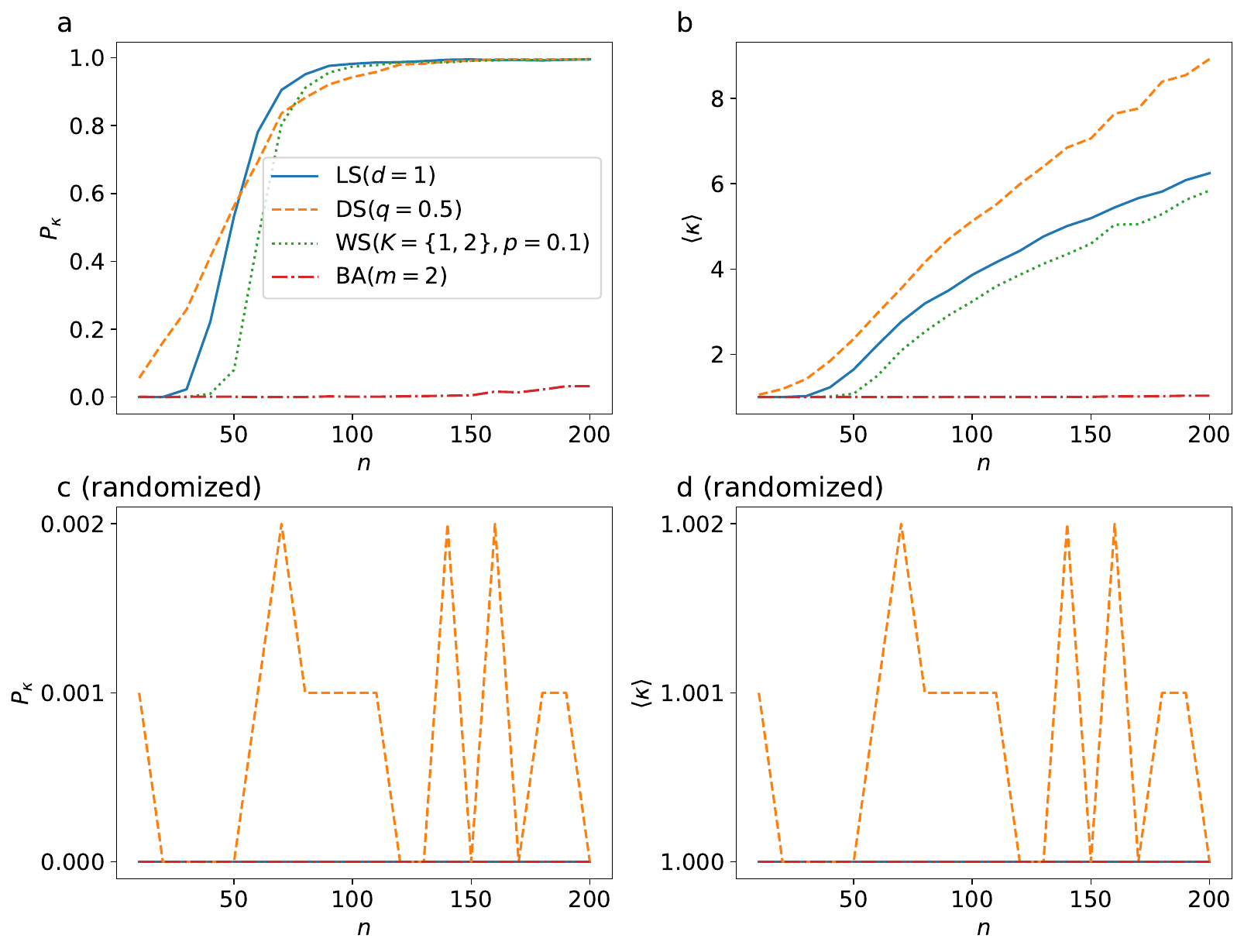}
\caption{a) Probability of having 2 or more communities and b) mean number of communities as a function of the number of nodes, for networks generated with the generative models indicated in the legend. c,d) Same data for the randomized networks preserving the nodes degrees. Communities were inferred with the stochastic block model with degree sequence correction.}
\label{fig_vs_n}
\end{figure}

\section{Network generative models}
\label{sec_models}

My hypothesis is that network communities can emerge as a consequence of local rules of evolution. To support this hypothesis I will consider some models with local network evolution rules and some without as negative examples.

\subsection{Local search}

I will start with the local search to depth $d$ model, ${\rm LS}(n, d)$. It is a version of the recursive search on a graph, where new links are created to visited nodes \cite{vazquez01rs}. {\em Initial condition:} The network is started with two connected nodes. {\em Evolution rule:} A new node is added and a $d$-steps random walk is performed from a randomly selected node in the current network. The new node is connected to all visited nodes. This model has preferential attachment because the probability that a node is visited, beyond the entry node, is proportional to the current nodes degrees. Consequently it generate networks with a power law degree distribution. The ${\rm LS}(n, d)$ networks have a high clustering coefficient as well. At least 1 triangle, between the entry point and next neighbor visited, is formed at every node addition.

Figure \ref{fig_ls1_example} shows an instance of ${\rm LS}(n=50, d=1)$. The stochastic block model with degree sequence correction detected two communities that we can corroborate from a visual inspection. However, there is no community structure encoded in the model evolution rules. They appeared by chance. In fact, some ${\rm LS}(n=50, d=1)$ instances have only 1 community containing all nodes. What fraction of all instances satisfy $c\geq 2$ depends on the network size (Fig. \ref{fig_vs_n}a, solid line). The probability $P_ \kappa = {\rm Prob}( \kappa\geq2)$ of getting an instance with 2 or more communities increases from zero at $n=10$ to almost 1 for $n>100$. The average number of communities increases monotonically from 1 beyond $n=25$  (Fig. \ref{fig_vs_n}b, solid line).  $P_\kappa = 1$ persists up to the largest network size tested $n = 100, 000$, where $\langle \kappa \rangle \approx 197$, based on stats for 10 network instances. This data suggests that for $n>100$ any instance of the ${\rm LS}(n, d=1)$ model has a community structure with about certainty.

Of note, I rewired the ${\rm LS}(n, d=1)$ networks preserving the nodes degrees. For these random graphs $P_ \kappa=0$ and $\langle\kappa\rangle =1$ (Fig. \ref{fig_vs_n}c,d, solid line).  They do not have a community structure.  This behavior persists up to the largest network size tested $n = 100, 000$. 

\begin{figure}
\includegraphics[width=3.3in]{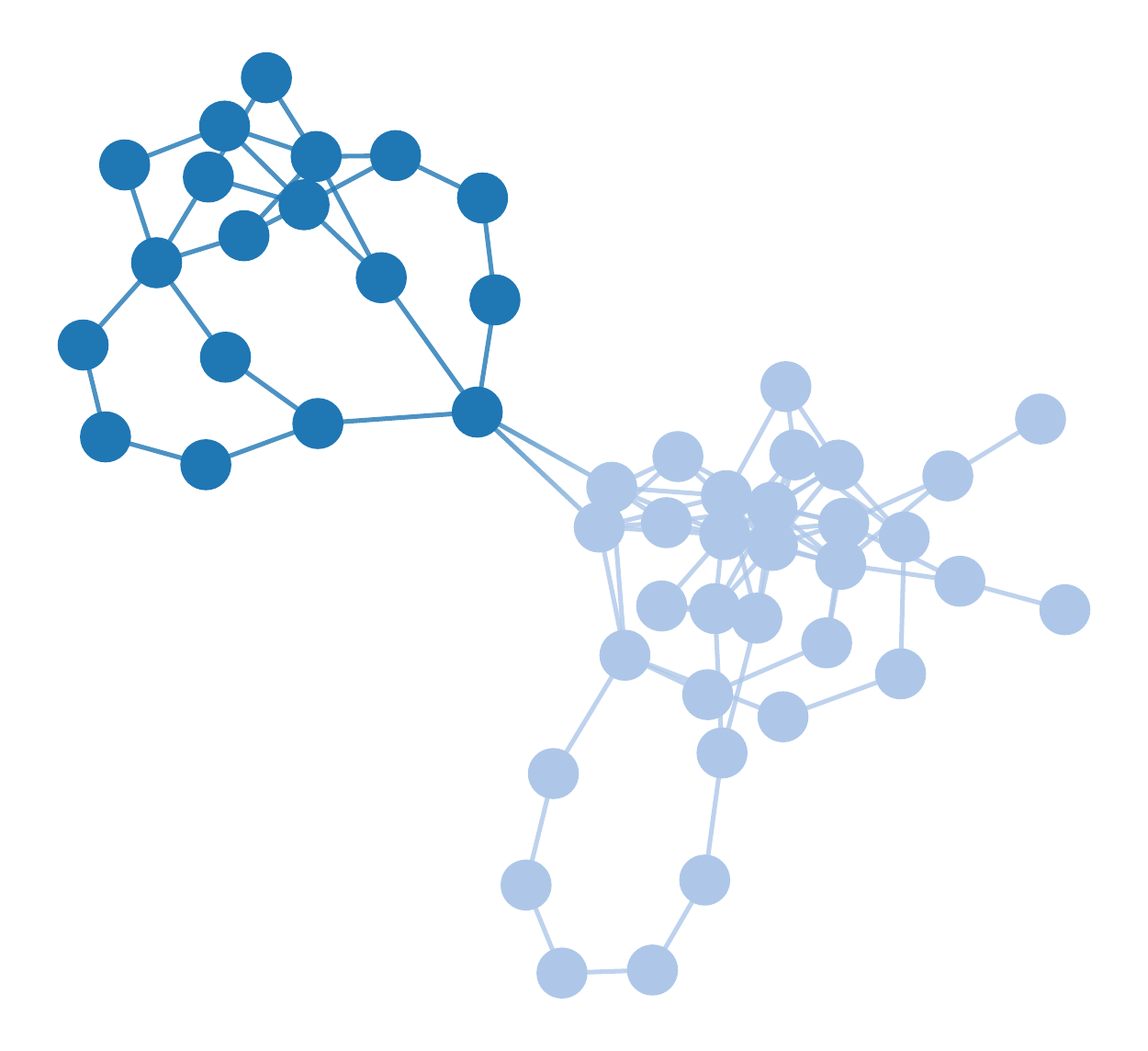}
\caption{An instance of DS$(n=50,\ q=0.3)$. The coloring represents the community structure inferred with the stochastic block model with degree sequence correction.}
\label{fig_ds0.3_example}
\end{figure}

\subsection{Duplication-split model}

The second local rule I study is the network evolution by node duplication or copying \cite{vazquez03dup, chung03, pastor-satorras03, copying_krapivsky_2005, vazquez_activity_2023}. In these generative models randomly selected nodes are duplicated, creating a new node with links to all neighbors of the parent node. The duplication rule induces a preferential attachment, because the probability that a neighbor of a node is duplicated is proportional to how many neighbors the node has, the node degree. However, other network properties can and do depend strongly on the choice of second rule. For example, the duplication rule does not make triangles, but triangles can be created by connecting the duplicates \cite{vazquez03dup}. I have chosen an undirected variant of the duplication-split model introduced in Ref. \cite{vazquez_activity_2023}, because it does not generate triangles.

The undirected duplication split-model with duplication rate $q$, ${\rm DS}(n, q)$ is defined as follows. {\em Initial condition:} The network is started with two connected nodes. {\em Evolution rule:} A new node $i$ is added to the network and a node in the current network is selected at random, node $j$. With probability $q$, $i$ becomes a duplicate of $j$ with links from $i$ to all neighbors of $j$. Otherwise, a link between $j$ and a randomly selected neighbor of $j$, node $k$, is split. The edge $(j, k)$ is removed and new edges $(i, j)$ and $(i,k)$ are created.

Figure \ref{fig_ds0.3_example} shows an instance of ${\rm DS}(n=50, q=0.3)$. The stochastic block model with degree sequence correction identified two communities. As I mentioned, the duplication rule does not make triangles and the split rule breaks triangles if they would exist. You can confirm that by looking and the example in Fig. \ref{fig_ds0.3_example}. It does not contain triangles. This observation indicates that a local structure is sufficient and that a high density of triangles is not a necessary condition.

The plot $P_ \kappa$ vs $n$ for the ${\rm DS}(n, q=0.5)$ networks is shown in Fig. \ref{fig_vs_n}, dashed line. It exhibits a similar pattern to what observed for ${\rm LS}(n, d=1)$. $P_\kappa$ increases from near 0 at $n=10$ to almost 1 for $n>150$. The average number of communities increases monotonically from 1 beyond $n=10$  (Fig. \ref{fig_vs_n}b, dashed line). $P_\kappa = 1$ persists up to the largest network size tested $n = 100, 000$, where $\langle \kappa \rangle \approx 343$, based on stats for 10 network instances. This is an example where communities appear even though the networks have a low clustering coefficient.

Of note, I rewired the ${\rm DS}(n, q=0.5)$ networks preserving the nodes degrees. For these random graphs $P_ \kappa\approx 0$ and $\langle\kappa\rangle\approx 1$ (Fig. \ref{fig_vs_n}c,d, dashed line). They do not have a community structure. That said, 1 or 2 out of the 1000 instances have two or more communities. Those can be interpreted as statistical glitches resulting from the stochastic nature of the method.  This behavior persists up to the largest network size tested $n = 100, 000$, where 1 out 10 network instances tested have more than 2 communities. 

\begin{figure}
\includegraphics[width=3.3in]{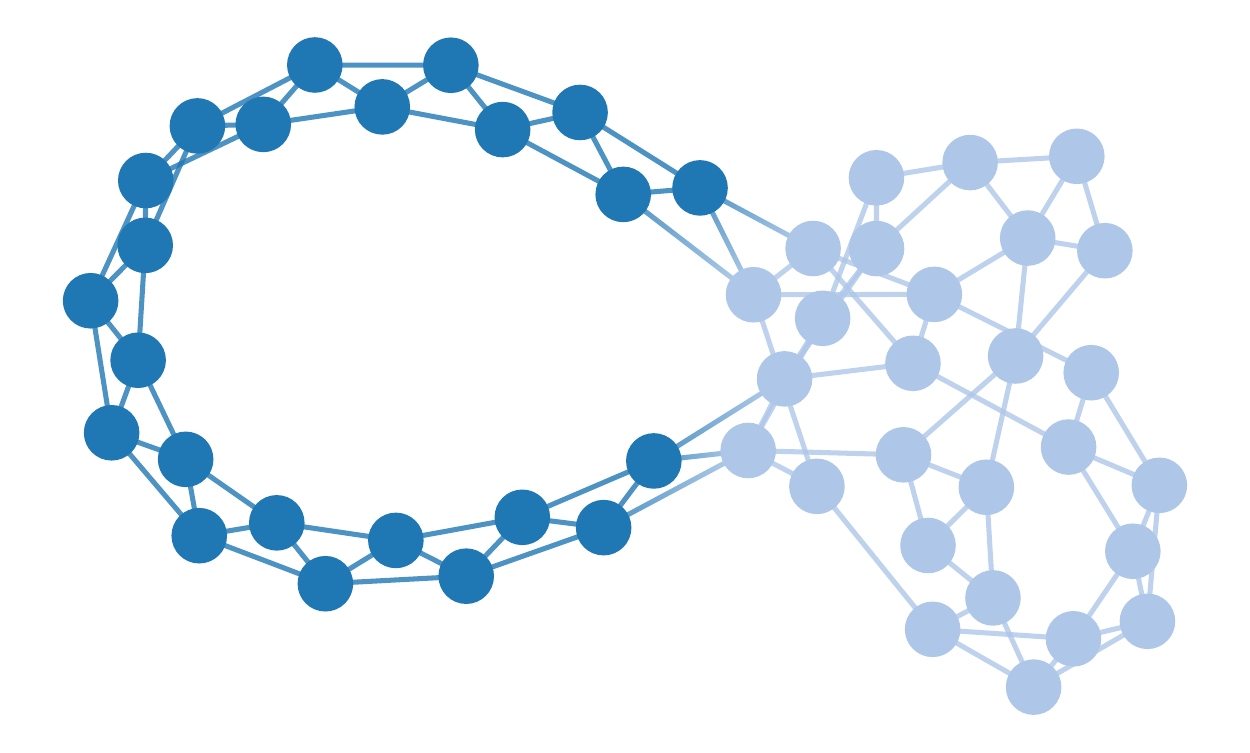}
\caption{An instance of WS$(n=50,\ K=\{1,2\},\ p=0.1)$. The coloring represents the community structure inferred with the stochastic block model with degree sequence correction.}
\label{fig_ws_example}
\end{figure}

\subsection{Watts-Strogatz model}
\label{sec:wats-strogatz}

The third model is a variation of the Watts-Strogatz model with neighbors set $K$ and rewiring rate $p$, ${\rm WS}(n, K, p)$ \cite{watts98}. The set $K$ represents the neighbors that are initially connected to a node. For example, $K=\{1,2\}$ means connected to the first and second neighbors to the right. The network is started with $i=1,\ldots, n$ nodes arranged in a ring. Each node $i$ is connected to the neighbor $i+s$ with probability $p$, otherwise to a randomly selected node, for all $s\in K$. The ${\rm WS}(n, K, p)$ networks have a local structure by construction, provided $p<1$. The original intend of Watts and Strogatz was to generate networks with a high clustering coefficient. In contrast, the random rewiring destroys the local order and therefore should work against the formation of communities.

Figure \ref{fig_ws_example} shows an instance of ${\rm WS}(n=50, K=\{1,2\},\ p=0.1)$. This instance has two communities that are verified by visual inspection. The communities retain part of the starting ring arrangement, giving them a different look than the previous examples.

The $P_ \kappa$ vs $n$ plot for the ${\rm WS}(n,\ K=\{1,2\},\ p=0.1)$ networks is shown in Fig. \ref{fig_vs_n}a, dotted line. $P_\kappa$ exhibits a similar pattern to what observed for ${\rm LS}(n, d=1)$ and ${\rm DS}(n, q=0.3)$. It increases from near 0 at $n=50$ to almost 1 for $n>100$. The average number of communities increases monotonically from 1 beyond $n=10$  (Fig. \ref{fig_vs_n}b, dotted line). $P_\kappa = 1$ persists up to the largest network size tested $n = 100, 000$, where $\langle \kappa \rangle \approx 202$, based on stats for 10 network instances. 

Of note, I rewired the ${\rm WS}(n, K, p=0.1)$ networks preserving the nodes degrees. For these random graphs $P_ \kappa=0$ and $\langle\kappa\rangle =1$ (Fig. \ref{fig_vs_n}c,d, dotted line). They do not have a community structure.This behavior persists up to the largest network size tested $n = 100, 000$.

\begin{figure}
\includegraphics[width=2in]{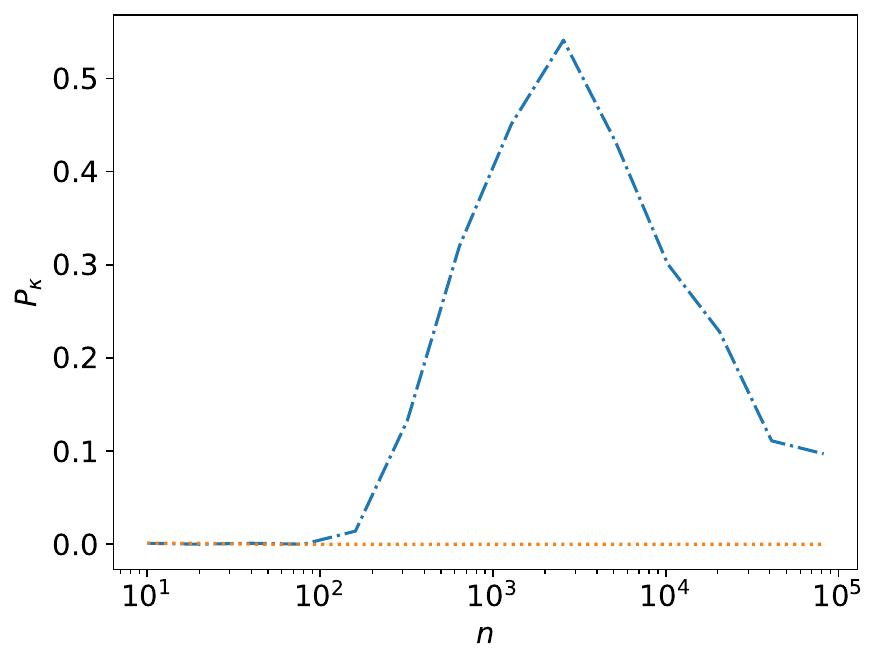}
\caption{Probability of having 2 or more communities as a function of the number of nodes, for networks generated with ${\rm BA}(n, m=2)$. Communities were inferred with the stochastic block model with degree sequence correction.}
\label{fig_ba_stats}
\end{figure}

\subsection{Barab\'asi-Albert model}
\label{sec:ba}

My fourth example is the Barab\'asi-Albert model with $m$ new links per added node, ${\rm BA}(n, m)$ \cite{barabasi99}, the canonical model of preferential attachment. {\em Initial condition:} The network is started with a fully connected graph on $m+1$ nodes. {\em Evolution rule:} A new node $i$ is added to the network and connected to $m$ nodes in the network, where $n$ is the current number of nodes in the network. The nodes are selected with a probability proportional to their current degree. I note there is no locality build in the ${\rm BA}(n, m)$ evolution rules. When a new node is added its $m$ neighbors are independently selected among all current nodes in the network.

In contrast to the generative models with local rules, the ${\rm BA}(n, m=2)$ model has $P_ \kappa = 0$ up to almost $n=150$ and increases slightly above 0 between $n=150$ and 200 (Fig. \ref{fig_vs_n}, dashed-dotted line). I have extended the analysis for larger $n$ values (Fig. \ref{fig_ba_stats}). $P_ \kappa$ reaches a maximum of about 0.5 for $n\sim 1000$ and from then on it decreases with increasing $n$. I have not found any explanation for this behavior. It could be rooted on the age structure of the ${\rm BA}(n, m)$ model where nodes added earlier have on the average a large degree than recently added nodes. The locality is induced by the age sequence. While this observation remains to be explained, it is evident the ${\rm BA}(n, m)$ instances do not have a community structure with almost certainty. The ${\rm BA}(n, m)$ networks have some structure, but not enough to warranty the appearance of communities.

\section{Other clustering methods}
\label{sec:other}

The data reported so far was obtained using the stochastic block model with degree sequence correction. Here I consider other community detection methods.

\begin{figure}
\includegraphics[width=3.4in]{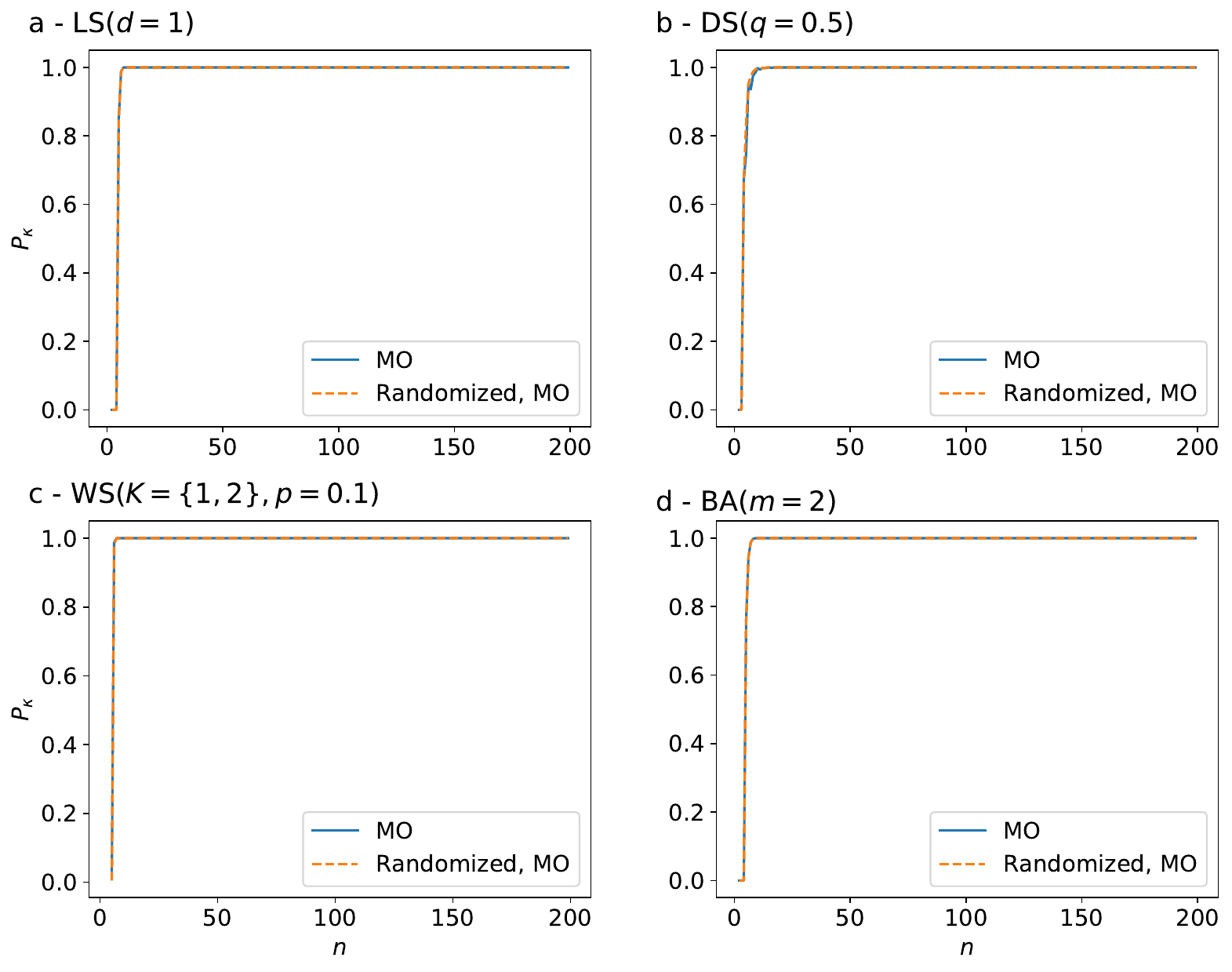}
\caption{Probability of having 2 or more communities ($P_\kappa$) for networks generated with different models and their randomized versions, when communities are inferred with modularity optimization (MO).}
\label{fig_vs_n_modularity}
\end{figure}

\subsection{Modularity optimization}

Modularity is typically used to find network communities \cite{newman_communities_2006}. It can be optimized using different methods to find the partition in network communities that maximizes the modularity \cite{communities_blondel_2008}. Figure \ref{fig_vs_n_modularity} reports the fraction of instances with at least 2 communities ($P_ \kappa$) when using the modularity optimization method. Based on modularity optimization $P_ \kappa=1$ for $n\geq10$ for all generative models investigated (Fig. \ref{fig_vs_n_modularity}a-d). And the same is true for the randomized versions of those networks instances. In fact, since $P_\kappa = 1$ at such small values of $n$ and for small $n$ the networks have little room for randomization, the plots of $P_ \kappa$ vs $n$ of the original and randomized networks overlap. These results agree with previous observations by Raddicchi {\em et al}, reporting that modularity optimization predicts network communities for random networks with given degree sequences \cite{radicchi_communities_2004}.

\begin{figure}
\includegraphics[width=3.4in]{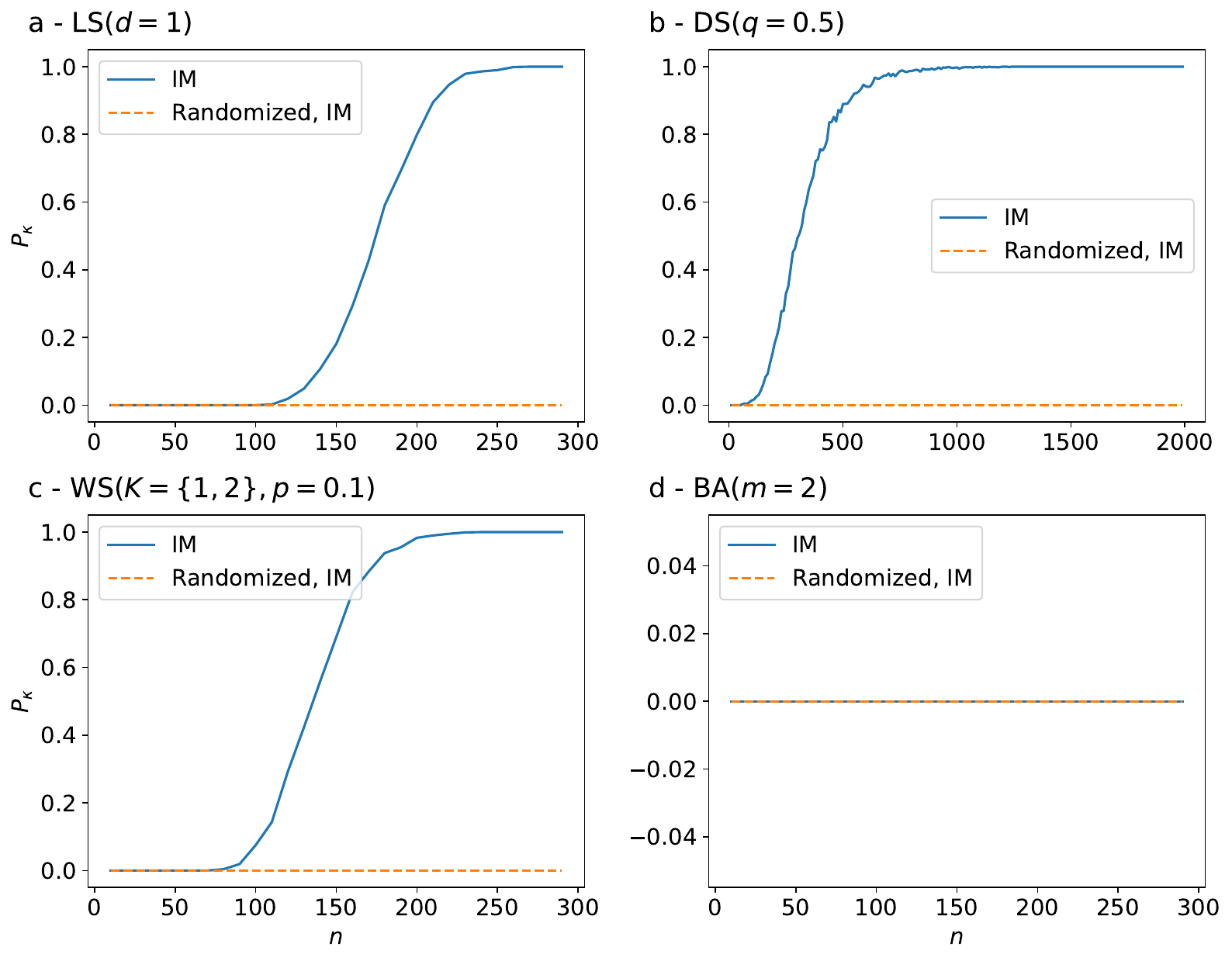}
\caption{Probability of having 2 or more communities ($P_\kappa$) for networks generated with different models  and their randomized versions, when communities are inferred using Infomap (IM) with regularization.}
\label{fig_vs_n_infomap}
\end{figure}

\subsection{Infomap}

Next I test the  infomap method \cite{mapequation2025software}. Infomap minimizes the description length of random walkers on the network \cite{communities_rosvall_2008}. Regularization increases the sensitivity of Infomap \cite{infomap_smiljani_2020}, correctly classifying random networks as having no communities. Therefore, I have set the flag \verb| regularized=True| and \verb| regularization_strength=1.0 |. The infomap method makes a distinction between the networks generated by local rules and their randomized versions preserving the degree distribution  (Fig. \ref{fig_vs_n_infomap}a-c, solid vs dashed lines). For the networks generated by local rules, $P_ \kappa$ reaches 1 at some finite $n$. For the randomized version of those networks Infomap predicts a single community confirming the high sensitivity of this method.  This evidence supports the data obtained with the SBM method, corroborating that local rules promote the emergence of network communities. For the Barab\'asi-Albert model, the Infomap method flags all networks instances as having a single community  (Fig. \ref{fig_vs_n_infomap}d), which was not the case with the SBM method (\ref{fig_ba_stats}).

For the networks generated by local rules, the Infomap method requires larger network sizes to reach $P_ \kappa\approx1$. Since there is no gold standard, we cannot tell which methods is right. Instead, we can say that these are different definitions of networks communities and that the emergence of network communities depend on how they are defined.

\begin{figure}
\includegraphics[width=3.4in]{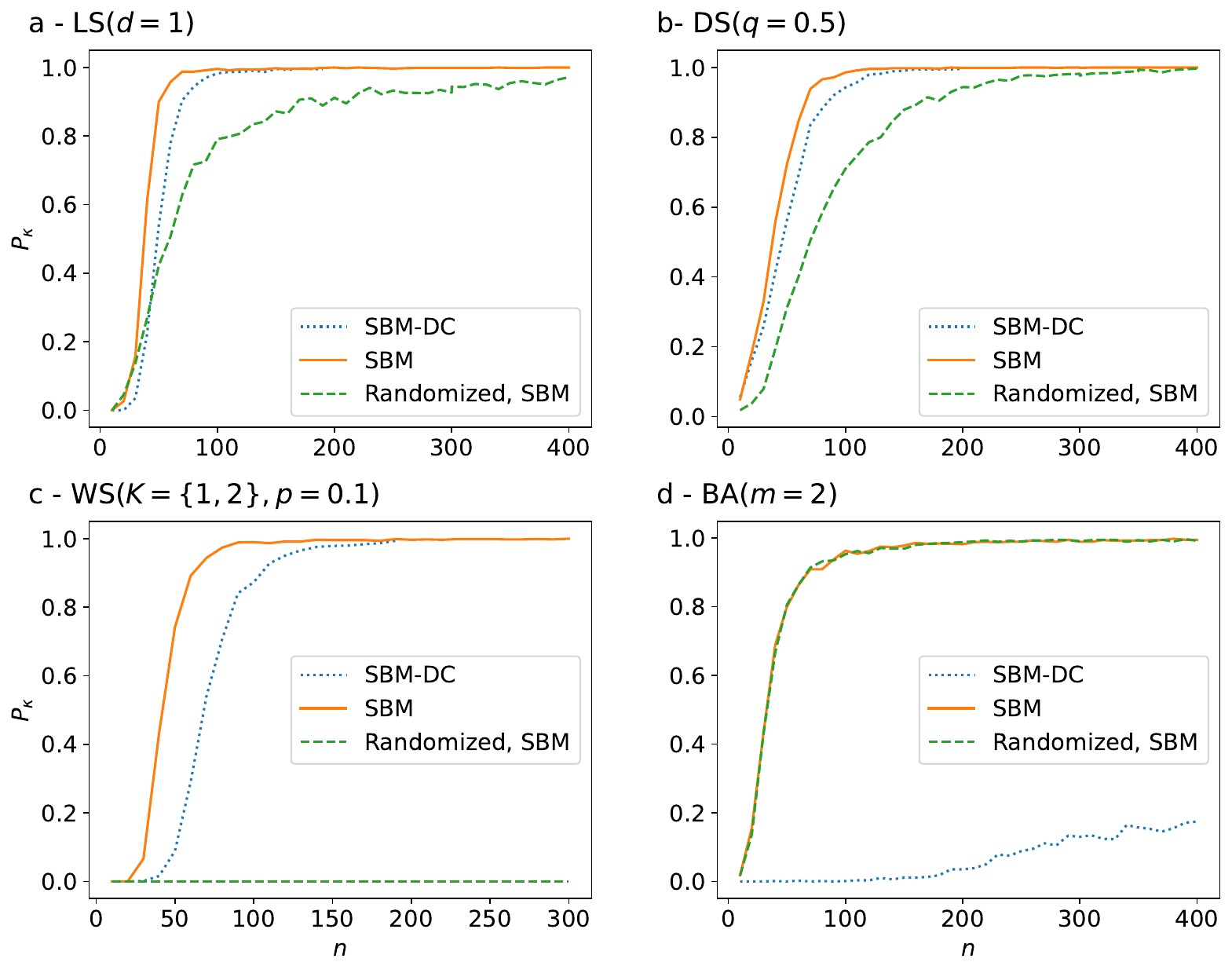}
\caption{Probability of having 2 or more communities ($P_\kappa$) for networks generated with different models  and their randomized versions, when communities are inferred with the stochastic block model with (SBM-DC) and without the degree sequence correction (SBM).}
\label{fig_vs_n_blocks-nodeg}
\end{figure}

\subsection{Stochastic block model without degree correction}

Modularity optimization fails to distinguish between networks with a local structure and their randomized versions preserving the nodes degrees. That may be rooted in the lack of regularization by this method. If a node has a small degree in cannot connect to many other nodes within a community. In contrast, a node with high can have many connections to nodes in a community, or be a bridge between multiple communities. To test this hypothesis I have repeated the analysis with the stochastic block model, now without correction for the degree sequence (Fig. \ref{fig_vs_n_blocks-nodeg}).

For the LS and DS networks, removing the degree sequence correction results in shift of the $P_\kappa$ vs $n$ plot to lower values of $n$ (Fig. \ref{fig_vs_n_blocks-nodeg}a,b, solid vs dotted line). More important, without degree correction the stochastic block model predicts the emergence of communities in the randomized networks as well, albeit at higher values of $n$ (Fig. \ref{fig_vs_n_blocks-nodeg}a,b dashed line). These observations demonstrate the importance of correcting for the degree sequence. Since this correction is missing in the modularity optimization, it fails in predicting the absence of communities in the randomized versions of the LS and DS networks.

For the WS networks, the picture is similar when comparing the stochastic block model with and without degree sequence correction. Removing the degree sequence correction shifts the $P_\kappa$ vs $n$ plot to lower values of $n$  (Fig. \ref{fig_vs_n_blocks-nodeg}c, solid and dotted line). In contrast with the LS and DS models, the randomized WS networks are characterized by $P_\kappa=0$  (Fig. \ref{fig_vs_n_blocks-nodeg}c, dashed line). That is explained by the lack of significant degree heterogeneity in the WS networks. In this case correcting for the degree sequence does not make a qualitative difference. 

Finally, for the BA model, the plot of $P_\kappa$ vs $n$ generated with the stochastic block model without degree sequence correction satures to 1 for $n$ larger than 200 (Fig. \ref{fig_vs_n_blocks-nodeg}d, solid line). That was not the case when correcting for the degree sequence (Fig. \ref{fig_vs_n_blocks-nodeg}d, dotted line). Furthermore, the plot of  $P_\kappa$ vs $n$ is almost identical for the randomized BA networks when not correcting for the degree sequence (Fig. \ref{fig_vs_n_blocks-nodeg}d, solid and dashed lines). The  BA networks have power degree distributions and here again correcting for the degree sequence is required to exclude the emergence of communities in the randomized networks.

Off note, similar results are obtained if we use the non-regularized version of Infomap (\verb| regularized=False|). Without regularization the Infomap method predicts the same qualitative behavior for the networks generated by local rules and their randomized versions.

This evidence consolidates the stochastic block model with degree sequence corrections and the Infomap method with regularization as the state of the art methods to detect networks communities. All the results reported below are obtained with these methods.

\section{Ramsey community number}
\label{sec_ramsey}

Inspired by these observations and to be more precise, I introduce the following definitions: 
\begin{definition}
Let $S_G$ be the set of all graphs,  $f_G(n): \rightarrow S_G$ a graph generative model on $n$ vertices, $f_ \kappa: S_G \rightarrow \mathbb{N}^+$ a graph community count function, $P_ \kappa: S_G\rightarrow [0,1]$ the communities likelihood function $P_ \kappa(f_G) = {\rm Prob}\{f_ \kappa[f_G(n)] \geq 2\}$ and $0<\epsilon<1$ an error rate. The Ramsey community number $r_ \kappa(f_G, f_ \kappa, \epsilon)$ is the minimum $n$ such that $P_ \kappa(f_G) \geq 1-\epsilon$.
\label{def:rc}
\end{definition}
\begin{definition}
The generative graph model on $n$ vertices $f_G(n)$ has the emergent communities property if $r_ \kappa(f_G, f_ \kappa, \epsilon)$ exist for all $0<\epsilon<1$.
\end{definition}

I estimate $r_ \kappa(\epsilon)$ by means of numerical simulations. The methodology proceeds as follows. First, identify an upper bound $n_{+}$ such that $P_ \kappa(f_G(n_{+}))\geq 1-\epsilon$. A simple approach is to start with some small value $n = n_{-}$ and duplicate $n$ until $P_ \kappa(f_G(n))\geq 1-\epsilon$ or $n>n_{\max}$, where $n_{\max}$ is a cutoff value to handle cases where $r_ \kappa$ does not exist. Second, conduct a binary search in the interval $[n_{-}, n_{+}]$ for the minimum value of $n$ satisfying $P_ \kappa(f_G(n))\geq 1-\epsilon$. 

In Fig. \ref{fig_vs_n}a we can see that $P_ \kappa = 1$ beyond a threshold $n$ for the three local models ${\rm LS}(n, d)$, ${\rm DS}(n, q)$ and ${\rm WS}(n, K, P)$. By means of numerical simulations I have confirmed that this is the case up to networks sizes of $n = 100,000$. They all satisfy the communities property. In contrast, from Fig. \ref{fig_ba_stats} it is evident that the BA model does not have the communities property: $r_ \kappa$ cannot be calculated for $P_ \kappa>0.6$ or equivalently $\epsilon<0.4$. In the following I will focus on $\epsilon = 0.05$, a 95\% confidence that a network instance has two or more communities. From now on I will write $r_ \kappa(f_G, f_\kappa)$ instead of $r_ \kappa(f_G, f_\kappa, \epsilon)$ since $\epsilon$ has been fixed.

\begin{table}[b]
\begin{tabular}{l|l|l}
Model & $r_ \kappa(SBM)$* & $r_ \kappa(IM)$* \\
\hline
${\rm LS}(d=1)$ & 81 & 222\\
${\rm LS}(d=2)$ & 77 & 212\\
${\rm LS}(d=3)$ & 87 & 297\\
${\rm DS}(q=0.3)$ & 239 & 1637\\
${\rm DS}(q=0.5)$ & 98 & 622\\
${\rm DS}(q=0.7)$ & 55 & 707\\
${\rm WS}(K=\{1\}, p)$ & $-$ & $-$\\
${\rm WS}(K=\{1,2\}, p=0.1)$ & 115 & 184\\
${\rm WS}(K=\{1,2\}, p=0.2)$ & 271 & 574\\
${\rm WS}(K=\{1,2\}, p=0.3)$ & 2377 & 2510\\
${\rm WS}(K=\{1,2,3\}, p=0.1)$ & 59 & 75\\
${\rm WS}(K=\{1,2,3\}, p=0.2)$ & 81 & 151\\
${\rm WS}(K=\{1,2,3\}, p=0.3)$ & 168 & 418\\
${\rm WS}(K=\{1,2,3\}, p=0.4)$ & 723 & 1783\\
${\rm BA}(m=2)$ & $-$ & $-$
\end{tabular}
\caption{Ramsey community numbers for different network models and $\epsilon=0.05$, when using the stochastic block model correcting for the degree sequence (SBM) or regularized Infomap (IM). $-$Does not have the emergent communities property. *An estimate, not an exact number, with variations in the last digit due to the stochastic nature of the network generative models and the communities detection method.}
\label{tab_rc_values}
\end{table}

The estimated $r_ \kappa(f_G, f_\kappa)$ for the models introduced above are reported in Table \ref{tab_rc_values}, for both  $f_\kappa={\rm SMB}$ and $f_\kappa={\rm Infomap}$. The ${\rm LS}(n, d)$, ${\rm DS}(n, q$) and ${\rm WS}(n, |K|\geq2, p)$ models all have finite $r_ \kappa$ values. These models exhibit the emergent communities property. In contrast, ${\rm WS}(n, K=\{2\}, p)$ and ${\rm BA}(n, m=2)$ do not exhibit the emergent communities property in numerical tests up to a graph size of $n=40,960$ nodes. This dichotomy indicates that having a local structure is a necessary condition to have the emergent communities property.

In the case of the ${\rm WS}(n, K, p)$ networks there is an evident increase of $r_\kappa$ with increasing $p$ (Tab. \ref{tab_rc_values}). As the networks get more randomized it becomes harder for the communities to emerge. This agrees with the expectation that the randomized networks do not have emergent communities. For the ${\rm WS}(n, K=\{1,2\}, p=0.4)$ networks $P\kappa$ increases with increasing $n$, reaching $P_\kappa=0.4$ for the largest network size tested $n=81920$. It is not clear whether $r_\kappa$ is very large or whether $r_\kappa=\infty$ beyond some $p_c(K)$. The behavior for large $p$ remains to be investigated.

The $r_ \kappa(f_G, f_\kappa)$ trends with the model parameters are similar when using SMB or Infomap, but there are quantitative differences. As anticipated in its definition, $r_ \kappa(f_G, f_\kappa)$ depends on the choice of community detection method $f_\kappa$.

\begin{figure}
\includegraphics[width=2in]{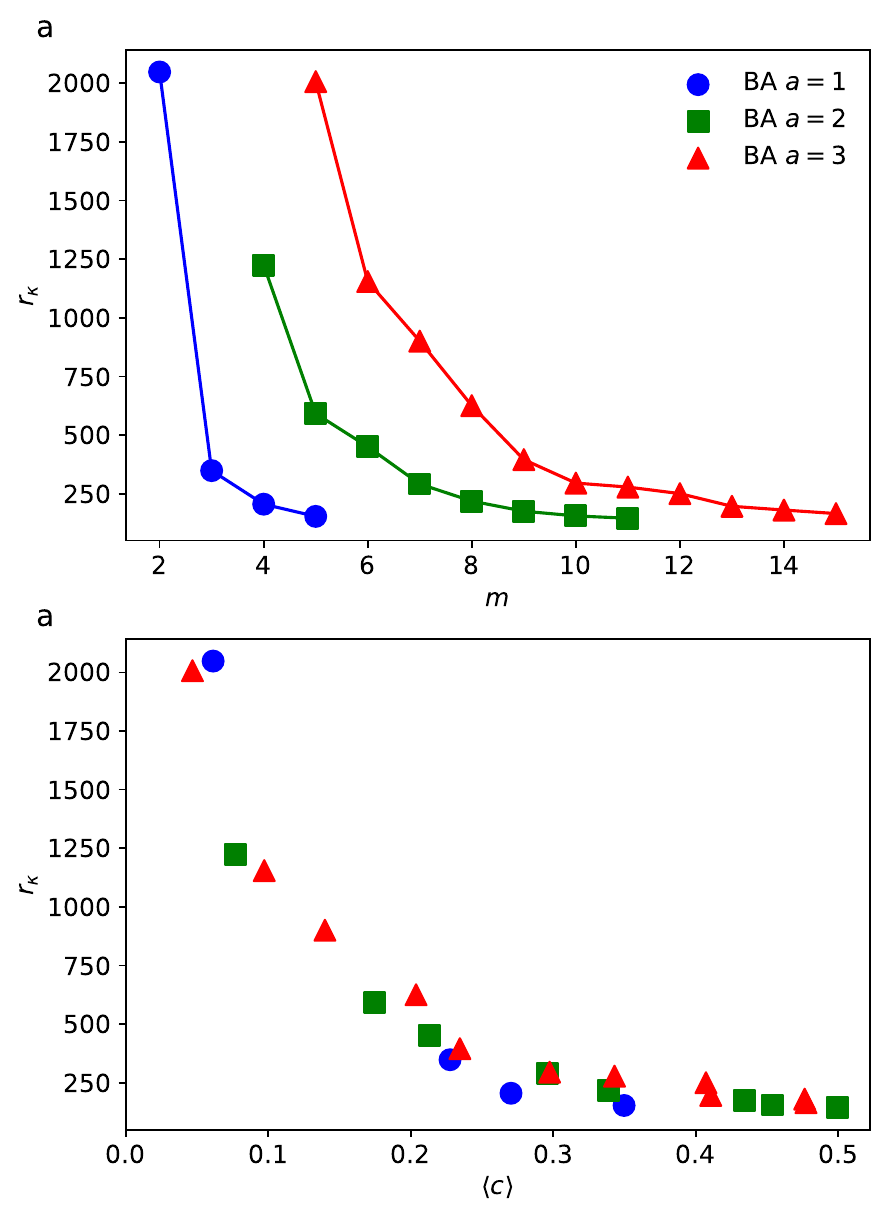}
\caption{Ramsey community number for networks generated with the ${\rm BA}(n, m, a)$ model. a) As a function of $m$ for different values of $a$. b) As a function of the clustering coefficient at $n=r_c$ for different values of $m$ and $a$. Communities were inferred with the stochastic block model with degree sequence correction.}
\label{fig_ba_a1_rc_vs_m}
\end{figure}

\section{Barabasi-Albert revisited}
\label{sec:ba_revisited}

The Barab\'asi-Albert model is an edge case between the generative models with local rules and their randomized versions. The class of models with Barab\'asi-Albert like rules includes variations like the preferential attachment model with attractiveness \cite{attractiveness_dorogovtsev_2000}, here denoted by ${\rm BA}(n, m, a)$. The initial condition and evolution rule is the same as in the original Barab\'asi-Albert model (Sec. \ref{sec:ba}), but there is an upgrade in the attachment rate. The nodes linked to the new node are selected with a probability $a + k_i / \sum_j k_j$, where $a$ represents a baseline attractiveness independent of the nodes degrees. This preferential attachment model generates networks with a power law degree distribution $p_k \sim k^{-\gamma(m,a)}$ with $\gamma(m,a) = 2 + a/m$ \cite{attractiveness_dorogovtsev_2000}. The case $a = m$ corresponds with the canonical Barab\'asi-Albert model with $\gamma(m, m) = 3$.

I have estimated the Ramsey community number of the ${\rm BA}(n, m, a)$ model for different values of $m$ and $a$ (Fig. \ref{fig_ba_a1_rc_vs_m}a). For a given value of $a$, $r_ \kappa$ decreases with increasing $m$ (decreasing $\gamma$). The $(m, a)$ values tested correspond with $\gamma(m,a)$ values less than 3. The largest $\gamma(m, a)$ value attained is $\gamma(5, 3) = 2 + 3/5 = 2.6$. All $(m, a)$ choices with $\gamma>2.6$ that I have tried show a similar behavior to that for the canonical Barab\'asi-Albert model (Fig. \ref{fig_ba_stats}).

Given the hypothesis that the local structures induce community structures, I have replotted the Ramsey community number as a function of the average clustering coefficient across nodes $\langle c\rangle$ (Fig . \ref{fig_ba_a1_rc_vs_m}b), where the clustering coefficient is calculated at $n = r_ \kappa$. With this rescaling, data for different choices of $(m, a)$ fall into the same overall trend, whereby $r_ \kappa$ is a monotonic decreasing function of $\langle c\rangle$. In turn, as $\langle c\rangle$ falls below 0.1, $r_ \kappa$ grows significantly, suggesting a diverging behavior for $\langle c\rangle \rightarrow 0$.

The divergence of $r_\kappa$ near $\langle c\rangle = 0$ brings some light in the lack of $r_\kappa$ value for the Barab\'asi-Albert model ${\rm BA}(n, m, m)$. The clustering coefficient of Barab\'asi-Albert networks decays with increasing the network size $n$ as $\langle c\rangle_{\rm BA} \sim (\ln n)^2/n$ \cite{clustering_klemm_2002}. A decrease in the density of cycles of larger length is expected as well. Therefore, as  the Barab\'asi-Albert networks get larger, they loosen their local structure and a lack of community structures is expected. That reasoning agrees with what observed in Fig. \ref{fig_ba_stats} to the right of the peak. The behavior to the left of the peak remains to be explained.

In summary, the ${\rm BA}(n, m, a)$ is a generative model without explicit local rules. The networks dynamics generates a high clustering coefficient leading to an implicit formation of local structures. In turn resulting in the emergence of communities, provided $\gamma(m, a)<3$.

\begin{figure}
\includegraphics[width=2in]{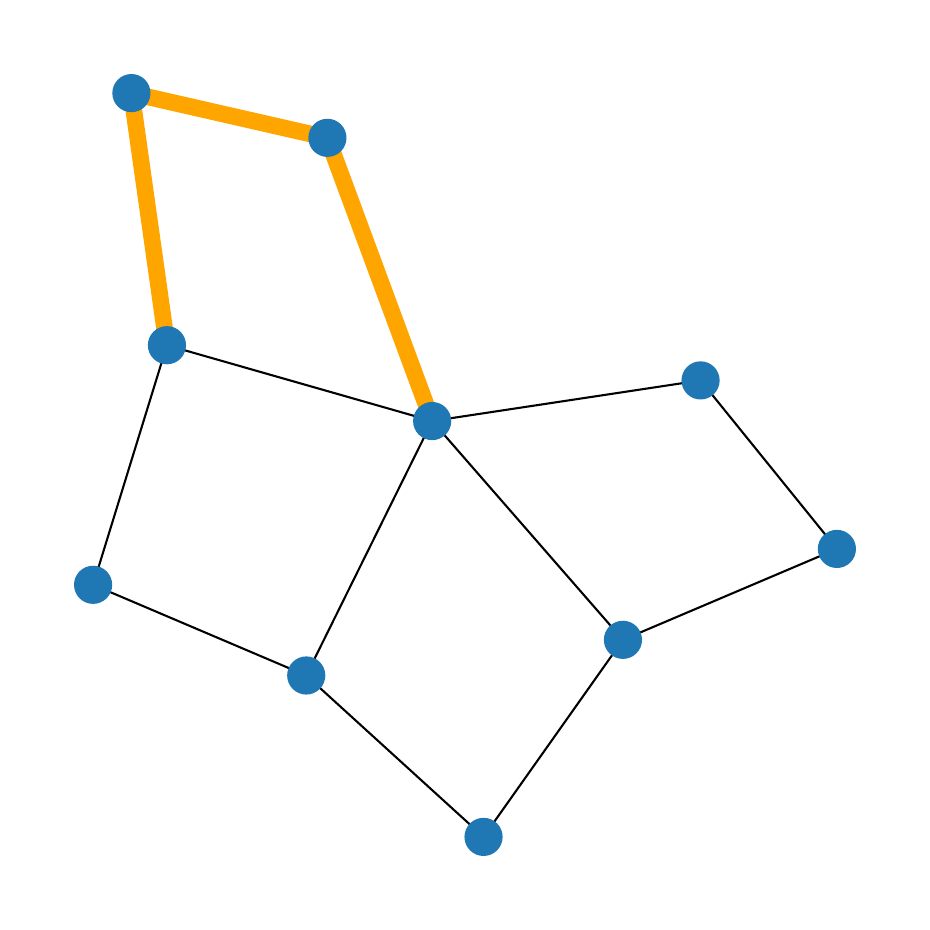}
\caption{An instance of ${\rm BB}(n=10, L=2)$. The orange/thick edges highlight the last chain of $L=2$ nodes added the to the network, forming a cycle of 4 nodes.}
\label{fig_bb_L2}
\end{figure}

\section{Dependency on cycle size}
\label{sec:cycles}

The thesis of this work is that local rules induce community structures. It is seeded on past work showing that triadic closure generate networks with communities \cite{growing_social_networks_jin01} and supplemented with the duplication-split model with minimum cycle size 4. To provide an example with variable cycle length I introduce the bubble model ${\rm BB}(n, L)$ defined as follows. {\em Initial condition:} The network is started with two connected nodes. {\em Evolution rule:} A chain of $L$ new nodes is added to the network and the two nodes at the chain ends are attached to the end of an existing link selected at random, forming a cycle of length $L+2$ (Fig. \ref{fig_bb_L2}). This evolution rule induces preferential attachment because the chance that a node $i$ with degree $k_i$ increases its connectivity upon addition of a chain is $2k_i / \sum_j k_j$, independently of $L$. More important for the subject of this section, by connecting the chain of length $L$ to the ends of a link a cycle of length $L+2$ is created.

\begin{figure}
\includegraphics[width=2in]{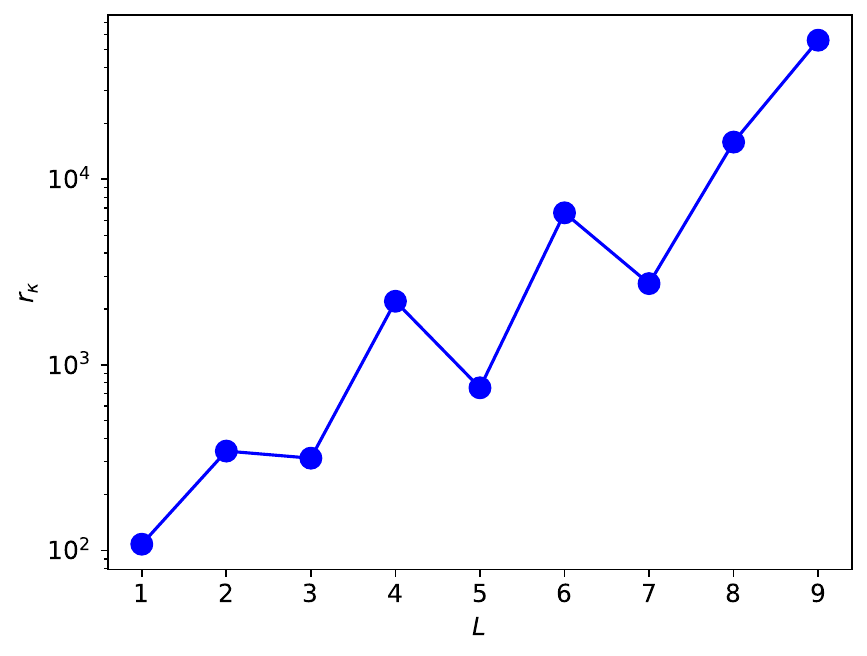}
\caption{Ramsey community number as a function of $L$ for networks generated with the ${\rm BB}(n, L)$ model. Communities were inferred with the stochastic block model with degree sequence correction.}
\label{fig_bb_rc_vs_L}
\end{figure}

Figure \ref{fig_bb_rc_vs_L} reports the Ramsey community number as a function of $L$ (cycles of length $L+2$). Overall there is an increasing trend of $r_ \kappa$ with increasing $L$.  On top of the increasing trend, there is an alternating pattern of $r_ \kappa$ vs $L$, up to $L=7$. What is the origin of this behavior remains to be understood. Nevertheless, the overall trend strengthen the hypothesis that local connections (small cycle lengths) induce the formation of network communities at lower network sizes.

\subsection{Breaking locality}

While the bubble model modulates the size of the minimum cycle it remains local. The ends of the chain are connected to the ends of a link. That is two nodes one step apart and, therefore, local. We can investigate the relevance of locality by relaxing the distance between the chosen nodes in the existing network. For example, the ends of the added chain can be connected to two distinct nodes that are at most a walk of $W$ steps apart, the ${\rm BB}(n, L, W)$ model. For $W=1$ we recover the bubble model above. For $W=2$ the connections are restricted to neighbors or nearest neighbors and so on. I have performed numerical simulations for ${\rm BB}(n, L=1, W)$ and determine the networks community structure using the SBM. I obtain the Ramsey community numbers $r_\kappa(L=1, W=1)=108$ and $r_\kappa(L=1, W=1)=1090$. For ${\rm BB}(n, L=1, W=3)$ there are no instances with communities up to the network sizes tested ($n=1280$). The latter is also true when the ends of the chain are connected to two distinct nodes chosen randomly and independently of the distance between them.  This data indicates that, for the bubble model, local means two steps apart. In agreement with common sense, communities will emerge if new connections appear between nearby nodes, those reinforcing local connections. In contrast, if new links can be formed between distant nodes it will be harder to observe the emergency of communities.

\section{Conclusions}
\label{sec_conclusions}

I have provided a precise definition of the {\em Ramsey community number} and the {\em emergent communities property}. The {\em Ramsey community number} allow us to determine how large a network structure needs to be to have a community structure with almost certainty. The {\em emergent communities property} states whether that number can actually be found for a given class of networks. These two concepts help to make precise statements about {\em emergence} in the context of network communities driven by the network topology alone.

When using the stochastic block model and Infomap, and by means of numerical simulations, I have demonstrated the emergence of network communities in generative models with explicit or implicit local structures, but not in their randomized versions preserving the degree sequence.

In a wider perspective, local network dynamics can explain the properties of real networks such as preferential attachment, power law degree distribution, high clustering coefficient, small-world and degree correlations \cite{vazquez03local}. Based on this work they can explain the emergence of network communities as well.

\bibliographystyle{apsrev4-1}


\input{ramsey_communities_v5.bbl}

\end{document}

%% file: ramsey_communities_v5.bbl
%

%% file: ramsey_communities_v5.bbl
\begin{thebibliography}{32}%
\makeatletter
\providecommand \@ifxundefined [1]{%
 \@ifx{#1\undefined}
}%
\providecommand \@ifnum [1]{%
 \ifnum #1\expandafter \@firstoftwo
 \else \expandafter \@secondoftwo
 \fi
}%
\providecommand \@ifx [1]{%
 \ifx #1\expandafter \@firstoftwo
 \else \expandafter \@secondoftwo
 \fi
}%
\providecommand \natexlab [1]{#1}%
\providecommand \enquote  [1]{``#1''}%
\providecommand \bibnamefont  [1]{#1}%
\providecommand \bibfnamefont [1]{#1}%
\providecommand \citenamefont [1]{#1}%
\providecommand \href@noop [0]{\@secondoftwo}%
\providecommand \href [0]{\begingroup \@sanitize@url \@href}%
\providecommand \@href[1]{\@@startlink{#1}\@@href}%
\providecommand \@@href[1]{\endgroup#1\@@endlink}%
\providecommand \@sanitize@url [0]{\catcode `\\12\catcode `\$12\catcode
  `\&12\catcode `\#12\catcode `\^12\catcode `\_12\catcode `\%12\relax}%
\providecommand \@@startlink[1]{}%
\providecommand \@@endlink[0]{}%
\providecommand \url  [0]{\begingroup\@sanitize@url \@url }%
\providecommand \@url [1]{\endgroup\@href {#1}{\urlprefix }}%
\providecommand \urlprefix  [0]{URL }%
\providecommand \Eprint [0]{\href }%
\providecommand \doibase [0]{http://dx.doi.org/}%
\providecommand \selectlanguage [0]{\@gobble}%
\providecommand \bibinfo  [0]{\@secondoftwo}%
\providecommand \bibfield  [0]{\@secondoftwo}%
\providecommand \translation [1]{[#1]}%
\providecommand \BibitemOpen [0]{}%
\providecommand \bibitemStop [0]{}%
\providecommand \bibitemNoStop [0]{.\EOS\space}%
\providecommand \EOS [0]{\spacefactor3000\relax}%
\providecommand \BibitemShut  [1]{\csname bibitem#1\endcsname}%
\let\auto@bib@innerbib\@empty
\bibitem [{\citenamefont {Newman}(2006)}]{newman_communities_2006}%
  \BibitemOpen
  \bibfield  {author} {\bibinfo {author} {\bibfnamefont {M.~E.~J.}\
  \bibnamefont {Newman}},\ }\href {\doibase 10.1073/pnas.0601602103} {\bibfield
   {journal} {\bibinfo  {journal} {Proceedings of the National Academy of
  Sciences}\ }\textbf {\bibinfo {volume} {103}},\ \bibinfo {pages} {8577}
  (\bibinfo {year} {2006})},\ \Eprint
  {http://arxiv.org/abs/https://www.pnas.org/doi/pdf/10.1073/pnas.0601602103}
  {https://www.pnas.org/doi/pdf/10.1073/pnas.0601602103} \BibitemShut {NoStop}%
\bibitem [{\citenamefont {Radicchi}\ \emph {et~al.}(2004)\citenamefont
  {Radicchi}, \citenamefont {Castellano}, \citenamefont {Cecconi},
  \citenamefont {Loreto},\ and\ \citenamefont
  {Parisi}}]{radicchi_communities_2004}%
  \BibitemOpen
  \bibfield  {author} {\bibinfo {author} {\bibfnamefont {F.}~\bibnamefont
  {Radicchi}}, \bibinfo {author} {\bibfnamefont {C.}~\bibnamefont
  {Castellano}}, \bibinfo {author} {\bibfnamefont {F.}~\bibnamefont {Cecconi}},
  \bibinfo {author} {\bibfnamefont {V.}~\bibnamefont {Loreto}}, \ and\ \bibinfo
  {author} {\bibfnamefont {D.}~\bibnamefont {Parisi}},\ }\href {\doibase
  10.1073/pnas.0400054101} {\bibfield  {journal} {\bibinfo  {journal}
  {Proceedings of the National Academy of Sciences}\ }\textbf {\bibinfo
  {volume} {101}},\ \bibinfo {pages} {2658} (\bibinfo {year} {2004})},\ \Eprint
  {http://arxiv.org/abs/https://www.pnas.org/doi/pdf/10.1073/pnas.0400054101}
  {https://www.pnas.org/doi/pdf/10.1073/pnas.0400054101} \BibitemShut {NoStop}%
\bibitem [{\citenamefont {Fortunato}\ and\ \citenamefont
  {Barthélemy}(2007)}]{fortunato_communities_2007}%
  \BibitemOpen
  \bibfield  {author} {\bibinfo {author} {\bibfnamefont {S.}~\bibnamefont
  {Fortunato}}\ and\ \bibinfo {author} {\bibfnamefont {M.}~\bibnamefont
  {Barthélemy}},\ }\href {\doibase 10.1073/pnas.0605965104} {\bibfield
  {journal} {\bibinfo  {journal} {Proceedings of the National Academy of
  Sciences}\ }\textbf {\bibinfo {volume} {104}},\ \bibinfo {pages} {36}
  (\bibinfo {year} {2007})},\ \Eprint
  {http://arxiv.org/abs/https://www.pnas.org/doi/pdf/10.1073/pnas.0605965104}
  {https://www.pnas.org/doi/pdf/10.1073/pnas.0605965104} \BibitemShut {NoStop}%
\bibitem [{\citenamefont {Hofman}\ and\ \citenamefont
  {Wiggins}(2008)}]{hofman_communities_2008}%
  \BibitemOpen
  \bibfield  {author} {\bibinfo {author} {\bibfnamefont {J.~M.}\ \bibnamefont
  {Hofman}}\ and\ \bibinfo {author} {\bibfnamefont {C.~H.}\ \bibnamefont
  {Wiggins}},\ }\href {\doibase 10.1103/PhysRevLett.100.258701} {\bibfield
  {journal} {\bibinfo  {journal} {Phys. Rev. Lett.}\ }\textbf {\bibinfo
  {volume} {100}},\ \bibinfo {pages} {258701} (\bibinfo {year}
  {2008})}\BibitemShut {NoStop}%
\bibitem [{\citenamefont {Karrer}\ and\ \citenamefont
  {Newman}(2011)}]{karrer_blockmodels_2011}%
  \BibitemOpen
  \bibfield  {author} {\bibinfo {author} {\bibfnamefont {B.}~\bibnamefont
  {Karrer}}\ and\ \bibinfo {author} {\bibfnamefont {M.~E.~J.}\ \bibnamefont
  {Newman}},\ }\href {\doibase 10.1103/PhysRevE.83.016107} {\bibfield
  {journal} {\bibinfo  {journal} {Phys. Rev. E}\ }\textbf {\bibinfo {volume}
  {83}},\ \bibinfo {pages} {016107} (\bibinfo {year} {2011})}\BibitemShut
  {NoStop}%
\bibitem [{\citenamefont
  {Peixoto}(2024)}]{peixoto2024networkreconstructionminimumdescription}%
  \BibitemOpen
  \bibfield  {author} {\bibinfo {author} {\bibfnamefont {T.~P.}\ \bibnamefont
  {Peixoto}},\ }\href {https://arxiv.org/abs/2405.01015} {\enquote {\bibinfo
  {title} {Network reconstruction via the minimum description length
  principle},}\ } (\bibinfo {year} {2024}),\ \Eprint
  {http://arxiv.org/abs/2405.01015} {arXiv:2405.01015 [stat.ML]} \BibitemShut
  {NoStop}%
\bibitem [{\citenamefont {Vazquez}(2009)}]{vazquez09hg}%
  \BibitemOpen
  \bibfield  {author} {\bibinfo {author} {\bibfnamefont {A.}~\bibnamefont
  {Vazquez}},\ }\href {\doibase 10.1088/1742-5468/2009/07/p07006} {\bibfield
  {journal} {\bibinfo  {journal} {J. Stat. Mech.: Theory Exp.}\ }\textbf
  {\bibinfo {volume} {2009}},\ \bibinfo {pages} {P07006} (\bibinfo {year}
  {2009})}\BibitemShut {NoStop}%
\bibitem [{\citenamefont {Contisciani}\ \emph {et~al.}(2022)\citenamefont
  {Contisciani}, \citenamefont {Battiston},\ and\ \citenamefont
  {De~Bacco}}]{contisciani_inference_2022}%
  \BibitemOpen
  \bibfield  {author} {\bibinfo {author} {\bibfnamefont {M.}~\bibnamefont
  {Contisciani}}, \bibinfo {author} {\bibfnamefont {F.}~\bibnamefont
  {Battiston}}, \ and\ \bibinfo {author} {\bibfnamefont {C.}~\bibnamefont
  {De~Bacco}},\ }\href {\doibase 10.1038/s41467-022-34714-7} {\bibfield
  {journal} {\bibinfo  {journal} {Nat Commun}\ }\textbf {\bibinfo {volume}
  {13}},\ \bibinfo {pages} {7229} (\bibinfo {year} {2022})},\ \bibinfo {note}
  {publisher: Nature Publishing Group}\BibitemShut {NoStop}%
\bibitem [{\citenamefont {Jin}\ \emph {et~al.}(2001)\citenamefont {Jin},
  \citenamefont {Girvan},\ and\ \citenamefont
  {Newman}}]{growing_social_networks_jin01}%
  \BibitemOpen
  \bibfield  {author} {\bibinfo {author} {\bibfnamefont {E.~M.}\ \bibnamefont
  {Jin}}, \bibinfo {author} {\bibfnamefont {M.}~\bibnamefont {Girvan}}, \ and\
  \bibinfo {author} {\bibfnamefont {M.~E.~J.}\ \bibnamefont {Newman}},\ }\href
  {\doibase 10.1103/PhysRevE.64.046132} {\bibfield  {journal} {\bibinfo
  {journal} {Phys. Rev. E}\ }\textbf {\bibinfo {volume} {64}},\ \bibinfo
  {pages} {046132} (\bibinfo {year} {2001})}\BibitemShut {NoStop}%
\bibitem [{\citenamefont {Solé}\ and\ \citenamefont
  {Valverde}(2007)}]{sole_spontaneous_2007}%
  \BibitemOpen
  \bibfield  {author} {\bibinfo {author} {\bibfnamefont {R.~V.}\ \bibnamefont
  {Solé}}\ and\ \bibinfo {author} {\bibfnamefont {S.}~\bibnamefont
  {Valverde}},\ }\href {\doibase 10.1098/rsif.2007.1108} {\bibfield  {journal}
  {\bibinfo  {journal} {Journal of The Royal Society Interface}\ }\textbf
  {\bibinfo {volume} {5}},\ \bibinfo {pages} {129} (\bibinfo {year} {2007})},\
  \bibinfo {note} {publisher: Royal Society}\BibitemShut {NoStop}%
\bibitem [{\citenamefont {Bianconi}\ \emph {et~al.}(2014)\citenamefont
  {Bianconi}, \citenamefont {Darst}, \citenamefont {Iacovacci},\ and\
  \citenamefont {Fortunato}}]{communities_bianconi_2014}%
  \BibitemOpen
  \bibfield  {author} {\bibinfo {author} {\bibfnamefont {G.}~\bibnamefont
  {Bianconi}}, \bibinfo {author} {\bibfnamefont {R.~K.}\ \bibnamefont {Darst}},
  \bibinfo {author} {\bibfnamefont {J.}~\bibnamefont {Iacovacci}}, \ and\
  \bibinfo {author} {\bibfnamefont {S.}~\bibnamefont {Fortunato}},\ }\href
  {\doibase 10.1103/PhysRevE.90.042806} {\bibfield  {journal} {\bibinfo
  {journal} {Phys. Rev. E}\ }\textbf {\bibinfo {volume} {90}},\ \bibinfo
  {pages} {042806} (\bibinfo {year} {2014})}\BibitemShut {NoStop}%
\bibitem [{\citenamefont {Wharrie}\ \emph {et~al.}(2019)\citenamefont
  {Wharrie}, \citenamefont {Azizi},\ and\ \citenamefont
  {Altmann}}]{communities_wharrie_2019}%
  \BibitemOpen
  \bibfield  {author} {\bibinfo {author} {\bibfnamefont {S.}~\bibnamefont
  {Wharrie}}, \bibinfo {author} {\bibfnamefont {L.}~\bibnamefont {Azizi}}, \
  and\ \bibinfo {author} {\bibfnamefont {E.~G.}\ \bibnamefont {Altmann}},\
  }\href {\doibase 10.1103/PhysRevE.100.022315} {\bibfield  {journal} {\bibinfo
   {journal} {Phys. Rev. E}\ }\textbf {\bibinfo {volume} {100}},\ \bibinfo
  {pages} {022315} (\bibinfo {year} {2019})}\BibitemShut {NoStop}%
\bibitem [{\citenamefont {Graham}\ \emph {et~al.}(1980)\citenamefont {Graham},
  \citenamefont {Rothschild},\ and\ \citenamefont
  {Spencer}}]{graham1980ramsey}%
  \BibitemOpen
  \bibfield  {author} {\bibinfo {author} {\bibfnamefont {R.}~\bibnamefont
  {Graham}}, \bibinfo {author} {\bibfnamefont {B.}~\bibnamefont {Rothschild}},
  \ and\ \bibinfo {author} {\bibfnamefont {J.}~\bibnamefont {Spencer}},\ }\href
  {https://books.google.de/books?id=S-juAAAAMAAJ} {\emph {\bibinfo {title}
  {Ramsey Theory}}},\ A Wiley-Interscience Publication\ (\bibinfo  {publisher}
  {Wiley},\ \bibinfo {year} {1980})\BibitemShut {NoStop}%
\bibitem [{\citenamefont {Vazquez}(2003)}]{vazquez03local}%
  \BibitemOpen
  \bibfield  {author} {\bibinfo {author} {\bibfnamefont {A.}~\bibnamefont
  {Vazquez}},\ }\href {\doibase 10.1103/PhysRevE.67.056104} {\bibfield
  {journal} {\bibinfo  {journal} {Phys. Rev. E}\ }\textbf {\bibinfo {volume}
  {67}},\ \bibinfo {pages} {056104} (\bibinfo {year} {2003})}\BibitemShut
  {NoStop}%
\bibitem [{\citenamefont {Peixoto}(2014)}]{peixoto_graph-tool_2014}%
  \BibitemOpen
  \bibfield  {author} {\bibinfo {author} {\bibfnamefont {T.~P.}\ \bibnamefont
  {Peixoto}},\ }\href {\doibase 10.6084/m9.figshare.1164194} {\bibfield
  {journal} {\bibinfo  {journal} {figshare}\ } (\bibinfo {year} {2014}),\
  10.6084/m9.figshare.1164194}\BibitemShut {NoStop}%
\bibitem [{\citenamefont {Peixoto}(2022)}]{triadic_peixoto_2022}%
  \BibitemOpen
  \bibfield  {author} {\bibinfo {author} {\bibfnamefont {T.~P.}\ \bibnamefont
  {Peixoto}},\ }\href {\doibase 10.1103/PhysRevX.12.011004} {\bibfield
  {journal} {\bibinfo  {journal} {Phys. Rev. X}\ }\textbf {\bibinfo {volume}
  {12}},\ \bibinfo {pages} {011004} (\bibinfo {year} {2022})}\BibitemShut
  {NoStop}%
\bibitem [{\citenamefont {Blondel}\ \emph {et~al.}(2008)\citenamefont
  {Blondel}, \citenamefont {Guillaume}, \citenamefont {Lambiotte},\ and\
  \citenamefont {Lefebvre}}]{communities_blondel_2008}%
  \BibitemOpen
  \bibfield  {author} {\bibinfo {author} {\bibfnamefont {V.~D.}\ \bibnamefont
  {Blondel}}, \bibinfo {author} {\bibfnamefont {J.-L.}\ \bibnamefont
  {Guillaume}}, \bibinfo {author} {\bibfnamefont {R.}~\bibnamefont
  {Lambiotte}}, \ and\ \bibinfo {author} {\bibfnamefont {E.}~\bibnamefont
  {Lefebvre}},\ }\href {\doibase 10.1088/1742-5468/2008/10/P10008} {\bibfield
  {journal} {\bibinfo  {journal} {Journal of Statistical Mechanics: Theory and
  Experiment}\ }\textbf {\bibinfo {volume} {2008}},\ \bibinfo {pages} {P10008}
  (\bibinfo {year} {2008})}\BibitemShut {NoStop}%
\bibitem [{\citenamefont {Hagberg}\ \emph {et~al.}(2008)\citenamefont
  {Hagberg}, \citenamefont {Swart},\ and\ \citenamefont {Schult}}]{networkx}%
  \BibitemOpen
  \bibfield  {author} {\bibinfo {author} {\bibfnamefont {A.}~\bibnamefont
  {Hagberg}}, \bibinfo {author} {\bibfnamefont {P.~J.}\ \bibnamefont {Swart}},
  \ and\ \bibinfo {author} {\bibfnamefont {D.~A.}\ \bibnamefont {Schult}}\
  }(\bibinfo {organization} {Los Alamos National Laboratory (LANL), Los Alamos,
  NM (United States)},\ \bibinfo {year} {2008})\BibitemShut {NoStop}%
\bibitem [{\citenamefont {Edler}\ \emph {et~al.}(2025)\citenamefont {Edler},
  \citenamefont {Holmgren},\ and\ \citenamefont
  {Rosvall}}]{mapequation2025software}%
  \BibitemOpen
  \bibfield  {author} {\bibinfo {author} {\bibfnamefont {D.}~\bibnamefont
  {Edler}}, \bibinfo {author} {\bibfnamefont {A.}~\bibnamefont {Holmgren}}, \
  and\ \bibinfo {author} {\bibfnamefont {M.}~\bibnamefont {Rosvall}},\
  }\href@noop {} {\enquote {\bibinfo {title} {{The MapEquation software
  package}},}\ }\bibinfo {howpublished} {\url{https://mapequation.org}}
  (\bibinfo {year} {2025})\BibitemShut {NoStop}%
\bibitem [{\citenamefont {Park}\ and\ \citenamefont
  {Newman}(2004)}]{configuration_method_park_2004}%
  \BibitemOpen
  \bibfield  {author} {\bibinfo {author} {\bibfnamefont {J.}~\bibnamefont
  {Park}}\ and\ \bibinfo {author} {\bibfnamefont {M.~E.~J.}\ \bibnamefont
  {Newman}},\ }\href {\doibase 10.1103/PhysRevE.70.066117} {\bibfield
  {journal} {\bibinfo  {journal} {Phys. Rev. E}\ }\textbf {\bibinfo {volume}
  {70}},\ \bibinfo {pages} {066117} (\bibinfo {year} {2004})}\BibitemShut
  {NoStop}%
\bibitem [{\citenamefont {Vazquez}(2001)}]{vazquez01rs}%
  \BibitemOpen
  \bibfield  {author} {\bibinfo {author} {\bibfnamefont {A.}~\bibnamefont
  {Vazquez}},\ }\href {\doibase 10.1209/epl/i2001-00259-y} {\bibfield
  {journal} {\bibinfo  {journal} {Europh. Lett.}\ }\textbf {\bibinfo {volume}
  {54}},\ \bibinfo {pages} {430} (\bibinfo {year} {2001})}\BibitemShut
  {NoStop}%
\bibitem [{\citenamefont {Vázquez}\ \emph {et~al.}(2003)\citenamefont
  {Vázquez}, \citenamefont {Flammini}, \citenamefont {Maritan},\ and\
  \citenamefont {Vespignani}}]{vazquez03dup}%
  \BibitemOpen
  \bibfield  {author} {\bibinfo {author} {\bibfnamefont {A.}~\bibnamefont
  {Vázquez}}, \bibinfo {author} {\bibfnamefont {A.}~\bibnamefont {Flammini}},
  \bibinfo {author} {\bibfnamefont {A.}~\bibnamefont {Maritan}}, \ and\
  \bibinfo {author} {\bibfnamefont {A.}~\bibnamefont {Vespignani}},\ }\href
  {\doibase 10.1159/000067642} {\bibfield  {journal} {\bibinfo  {journal}
  {Complexus}\ }\textbf {\bibinfo {volume} {1}},\ \bibinfo {pages} {38}
  (\bibinfo {year} {2003})}\BibitemShut {NoStop}%
\bibitem [{\citenamefont {Chung}\ \emph {et~al.}(2003)\citenamefont {Chung},
  \citenamefont {Lu}, \citenamefont {Dewey},\ and\ \citenamefont
  {Galas}}]{chung03}%
  \BibitemOpen
  \bibfield  {author} {\bibinfo {author} {\bibfnamefont {F.}~\bibnamefont
  {Chung}}, \bibinfo {author} {\bibfnamefont {L.}~\bibnamefont {Lu}}, \bibinfo
  {author} {\bibfnamefont {T.~G.}\ \bibnamefont {Dewey}}, \ and\ \bibinfo
  {author} {\bibfnamefont {D.~J.}\ \bibnamefont {Galas}},\ }\href {\doibase
  10.1089/106652703322539024} {\bibfield  {journal} {\bibinfo  {journal} {J.
  Comp. Biol.}\ }\textbf {\bibinfo {volume} {10}},\ \bibinfo {pages} {677}
  (\bibinfo {year} {2003})}\BibitemShut {NoStop}%
\bibitem [{\citenamefont {Pastor-Satorras}\ \emph {et~al.}(2003)\citenamefont
  {Pastor-Satorras}, \citenamefont {Smith},\ and\ \citenamefont
  {Solé}}]{pastor-satorras03}%
  \BibitemOpen
  \bibfield  {author} {\bibinfo {author} {\bibfnamefont {R.}~\bibnamefont
  {Pastor-Satorras}}, \bibinfo {author} {\bibfnamefont {E.}~\bibnamefont
  {Smith}}, \ and\ \bibinfo {author} {\bibfnamefont {R.~V.}\ \bibnamefont
  {Solé}},\ }\href {\doibase 10.1016/S0022-5193(03)00028-6} {\bibfield
  {journal} {\bibinfo  {journal} {J. Theor. Biol.}\ }\textbf {\bibinfo {volume}
  {222}},\ \bibinfo {pages} {199} (\bibinfo {year} {2003})}\BibitemShut
  {NoStop}%
\bibitem [{\citenamefont {Krapivsky}\ and\ \citenamefont
  {Redner}(2005)}]{copying_krapivsky_2005}%
  \BibitemOpen
  \bibfield  {author} {\bibinfo {author} {\bibfnamefont {P.~L.}\ \bibnamefont
  {Krapivsky}}\ and\ \bibinfo {author} {\bibfnamefont {S.}~\bibnamefont
  {Redner}},\ }\href {\doibase 10.1103/PhysRevE.71.036118} {\bibfield
  {journal} {\bibinfo  {journal} {Phys. Rev. E}\ }\textbf {\bibinfo {volume}
  {71}},\ \bibinfo {pages} {036118} (\bibinfo {year} {2005})}\BibitemShut
  {NoStop}%
\bibitem [{\citenamefont {Vazquez}\ \emph {et~al.}(2023)\citenamefont
  {Vazquez}, \citenamefont {Pozzana}, \citenamefont {Kalogridis},\ and\
  \citenamefont {Ellinas}}]{vazquez_activity_2023}%
  \BibitemOpen
  \bibfield  {author} {\bibinfo {author} {\bibfnamefont {A.}~\bibnamefont
  {Vazquez}}, \bibinfo {author} {\bibfnamefont {I.}~\bibnamefont {Pozzana}},
  \bibinfo {author} {\bibfnamefont {G.}~\bibnamefont {Kalogridis}}, \ and\
  \bibinfo {author} {\bibfnamefont {C.}~\bibnamefont {Ellinas}},\ }\href
  {\doibase 10.1038/s41598-022-27180-0} {\bibfield  {journal} {\bibinfo
  {journal} {Sci Rep}\ }\textbf {\bibinfo {volume} {13}},\ \bibinfo {pages}
  {509} (\bibinfo {year} {2023})}\BibitemShut {NoStop}%
\bibitem [{\citenamefont {Watts}\ and\ \citenamefont
  {Strogatz}(1998)}]{watts98}%
  \BibitemOpen
  \bibfield  {author} {\bibinfo {author} {\bibfnamefont {D.~J.}\ \bibnamefont
  {Watts}}\ and\ \bibinfo {author} {\bibfnamefont {S.~H.}\ \bibnamefont
  {Strogatz}},\ }\href@noop {} {\bibfield  {journal} {\bibinfo  {journal}
  {Nature}\ }\textbf {\bibinfo {volume} {393}},\ \bibinfo {pages} {440}
  (\bibinfo {year} {1998})}\BibitemShut {NoStop}%
\bibitem [{\citenamefont {Barabási}\ and\ \citenamefont
  {Albert}(1999)}]{barabasi99}%
  \BibitemOpen
  \bibfield  {author} {\bibinfo {author} {\bibfnamefont {A.-L.}\ \bibnamefont
  {Barabási}}\ and\ \bibinfo {author} {\bibfnamefont {R.}~\bibnamefont
  {Albert}},\ }\href {\doibase 10.1126/science.286.5439.509} {\bibfield
  {journal} {\bibinfo  {journal} {Science}\ }\textbf {\bibinfo {volume}
  {286}},\ \bibinfo {pages} {509} (\bibinfo {year} {1999})}\BibitemShut
  {NoStop}%
\bibitem [{\citenamefont {Rosvall}\ and\ \citenamefont
  {Bergstrom}(2008)}]{communities_rosvall_2008}%
  \BibitemOpen
  \bibfield  {author} {\bibinfo {author} {\bibfnamefont {M.}~\bibnamefont
  {Rosvall}}\ and\ \bibinfo {author} {\bibfnamefont {C.~T.}\ \bibnamefont
  {Bergstrom}},\ }\href {\doibase 10.1073/pnas.0706851105} {\bibfield
  {journal} {\bibinfo  {journal} {Proceedings of the National Academy of
  Sciences}\ }\textbf {\bibinfo {volume} {105}},\ \bibinfo {pages} {1118}
  (\bibinfo {year} {2008})},\ \Eprint
  {http://arxiv.org/abs/https://www.pnas.org/doi/pdf/10.1073/pnas.0706851105}
  {https://www.pnas.org/doi/pdf/10.1073/pnas.0706851105} \BibitemShut {NoStop}%
\bibitem [{\citenamefont {Smiljani\ifmmode~\acute{c}\else \'{c}\fi{}}\ \emph
  {et~al.}(2020)\citenamefont {Smiljani\ifmmode~\acute{c}\else \'{c}\fi{}},
  \citenamefont {Edler},\ and\ \citenamefont
  {Rosvall}}]{infomap_smiljani_2020}%
  \BibitemOpen
  \bibfield  {author} {\bibinfo {author} {\bibfnamefont {J.}~\bibnamefont
  {Smiljani\ifmmode~\acute{c}\else \'{c}\fi{}}}, \bibinfo {author}
  {\bibfnamefont {D.}~\bibnamefont {Edler}}, \ and\ \bibinfo {author}
  {\bibfnamefont {M.}~\bibnamefont {Rosvall}},\ }\href {\doibase
  10.1103/PhysRevE.102.012302} {\bibfield  {journal} {\bibinfo  {journal}
  {Phys. Rev. E}\ }\textbf {\bibinfo {volume} {102}},\ \bibinfo {pages}
  {012302} (\bibinfo {year} {2020})}\BibitemShut {NoStop}%
\bibitem [{\citenamefont {Dorogovtsev}\ \emph {et~al.}(2000)\citenamefont
  {Dorogovtsev}, \citenamefont {Mendes},\ and\ \citenamefont
  {Samukhin}}]{attractiveness_dorogovtsev_2000}%
  \BibitemOpen
  \bibfield  {author} {\bibinfo {author} {\bibfnamefont {S.~N.}\ \bibnamefont
  {Dorogovtsev}}, \bibinfo {author} {\bibfnamefont {J.~F.~F.}\ \bibnamefont
  {Mendes}}, \ and\ \bibinfo {author} {\bibfnamefont {A.~N.}\ \bibnamefont
  {Samukhin}},\ }\href {\doibase 10.1103/PhysRevLett.85.4633} {\bibfield
  {journal} {\bibinfo  {journal} {Phys. Rev. Lett.}\ }\textbf {\bibinfo
  {volume} {85}},\ \bibinfo {pages} {4633} (\bibinfo {year}
  {2000})}\BibitemShut {NoStop}%
\bibitem [{\citenamefont {Klemm}\ and\ \citenamefont
  {Egu\'{\i}luz}(2002)}]{clustering_klemm_2002}%
  \BibitemOpen
  \bibfield  {author} {\bibinfo {author} {\bibfnamefont {K.}~\bibnamefont
  {Klemm}}\ and\ \bibinfo {author} {\bibfnamefont {V.~M.}\ \bibnamefont
  {Egu\'{\i}luz}},\ }\href {\doibase 10.1103/PhysRevE.65.057102} {\bibfield
  {journal} {\bibinfo  {journal} {Phys. Rev. E}\ }\textbf {\bibinfo {volume}
  {65}},\ \bibinfo {pages} {057102} (\bibinfo {year} {2002})}\BibitemShut
  {NoStop}%
\end{thebibliography}
